# User modeling for point-of-interest recommendations in location-based social networks: the state-of-the-art

**Shudong Liu**

School of Information & Security Engineering, Zhongnan University of Economics & Law, Wuhan 430073, China

Correspondence should be addressed to Shudong Liu; bupt.mymeng@gmail.com

**Abstract**: The rapid growth of location-based services (LBSs) has greatly enriched people's urban lives and attracted millions of users in recent years. Location-based social networks (LBSNs) allow users to check-in at a physical location and share daily tips on points-of-interest (POIs) with their friends anytime and anywhere. Such check-in behavior can make daily real-life experiences spread quickly through the Internet. Moreover, such check-in data in LBSNs can be fully exploited to understand the basic laws of humans' daily movement and mobility. This paper focuses on reviewing the taxonomy of user modeling for POI recommendations through the data analysis of LBSNs. First, we briefly introduce the structure and data characteristics of LBSNs, then we present a formalization of user modeling for POI recommendations in LBSNs. Depending on which type of LBSNs data was fully utilized in user modeling approaches for POI recommendations, we divide user modeling algorithms into four categories: pure check-in data-based user modeling, geographical information-based user modeling, spatio-temporal information-based user modeling, and geo-social information-based user modeling. Finally, summarizing the existing works, we point out the future challenges and new directions in five possible aspects.

## 1、Introduction

The advanced information technologies that have resulted from the rapid growth of location-based services (LBSs) have greatly enriched people's urban lives. Location-based social networks (LBSNs) allow users to check-in and share their locations, tips, and experiences about points-of-interest (POIs) with their friends anytime and anywhere. For example, while having lunch at a restaurant, we may take photos of the dishes on the table and immediately share these photos with our friends via LBSNs. Such check-in behavior can make real-life daily experiences spread quickly over the Internet. Moreover, such check-in data of LBSNs can be fully exploited to understand the basic laws of human daily movement and mobility [1], which can be applied to recommendation systems and locat-ion-based services. Thus, location-based social media data services are attracting signifi cant attention from different commerce doma-ins, e.g., user profiling [1-3], recommendati-on systems [4,5], urban emergency event man-agement [6-9], urban planning [10] and mark-eting decisions [11].

User generated spatial-temporal data can be collected from LBSNs and can be widely used for understanding and modeling human mobility according to the following four aspects:

(1) Geographical feature

The spatial features of human movement as hidden in millions of check-in data has been exploited to understand human mobility. For example, people tend to move to nearby places and occasionally to distant places [2, 4], the former is short-ranged travel and is not affected by social network ties, which are periodic both spatially and temporally; the latter is long-distance travel and more influenced by social network ties [1].

(2) Temporal features.

the routines and habits of our daily lives, there are different probabilities for different locations at different hours of the day and different days of the week. The check-in data of LBSBs also reveals these results [3,5]. Most people go to work on the weekdays, their check-in behaviors often happen at noon or at night, and the locations they choose are close their workplaces or homes. On the weekends, most check-in behaviors happen in the morning or afternoon, and the locations are close to certain POIs (e.g. a marketplace, restaurant, museum, or scenic spot).

(3) Social features

First, many research studies [1, 12] show that people tend to visit close places more often than distant places, but they tend to visit distant places close to their friends' homes or those that are checked-in by their friends. These observations have been widely used for location recommendations in LBSNs [13-15]. Second, the spatial-temporal feature abstract -ted from check-in data has been exploited to infer social ties [16] and friend recommenda -tions [17-19].

(4) Integrated feature

As one type of global public data source about individual activity-related choices, the check-in data in LBSNs provides a new way to sense people's spatial and temporal preferences and infer their social ties. More -over, it always provides a new perspective from which urban structures and related socio -economic performances can be portrayed, street networks and POI popularity can be estimated [10, 20], intra-urban movement flows can be analyzed in urban areas [21, 22], urban major/emergency events can be identif -ied [6-9] and social-economic impacts of cultural investments can be detected [11].

Though several surveys on POI recomme -ndation have been published, few current studies present a formalization of user model -ing for POI recommendations in LBSNs and classify existing user modeling approaches based on which type of LBSNs data. This paper focuses on reviewing how we can efficiently make use of user-generated data to model POI recommendations in LBSNs. The contributions of this paper are as follows:

(1) Briefly introduce the structure and data characteristics of LBSNs. LBSNs can be abstracted into a three-layers + one-timeline framework. There are three types of data in LNSNs and six distinct characteristics of LBS Ns data.

(2) Considering the data characteristics of geographical and social data in LBSNs, we present a formalization of user modeling for POI recommendations in LBSNs.

(3) According to the type of LBSN data that is fully utilized in user modeling approach -es for POI recommendations, we divide user modeling algorithms into four categories: pure check-in data-based user modeling, geographi -cal information-based user modeling, spatio-temporal information-based user modeling and geo-social information based user mode -ling.

## 2、Characteristics of LBSNs

### 2.1 Structure of LBSNs

LBSNs are based on traditional online social networks and provide location-based services that allow users to check-in at physi -cal places and share location-related inform -ation with their friends. Meanwhile, LBSNs provide a new perspective for bridging the gap between the real and virtual worlds that allow users' real-life geographical activities to be disseminated on the Internet. The descriptive definition of LBSNs is given by Zheng et al. [23]. From this descriptive definition, LBSNs can be abstracted into a "3 + 1" framework [24], namely three layers and one timeline, as shown in Figure 1. The geographical layer is composed of users' historical check-ins, and the social layer is composed of users' friend -ships, while the content layer contains the media (i.e. photo, video, and text) that has been shared by users.

There are six types of relationship in LBSNs: location-location networks, user-user networks, media-media networks, user-loca -tion networks, user-media networks, and loca -tion-media networks. Traditionally, location-location networks, user-location networks, and location-media networks are the key research content and most user modeling for POI reco -mmendations in LBSNs are designed from the aspect of data mining and analyzing these three networks.

### 2.2 The data characteristics of LBSNs

There are three types of data in LBSNs: (1) user check-ins: the data records users' check-ins at different geographical locations at different times, (2) users' social relationships: the data records users' social relationships; (3) social activities: the data records the social activities where users participate at different geographical locations and at different times, or shared social media information. Except for users' social relationships, users' check-ins and social activities are the distinct properties of LBSNs data. More appropriately, they bridge the gaps between the real and virtual worlds in LBSNs.

In general, LBSNs' data characteristics can be summarized as follows:

(1) Multi-layer heterogeneous networks
As shown in Fig. 1, there are three different networks in LBSNs: check-in trajectory networks, social networks, and social media networks. The nodes and edges in the three networks are entirely disparate. Further-more, different networks also exist between any two of the above mentioned three net-works.

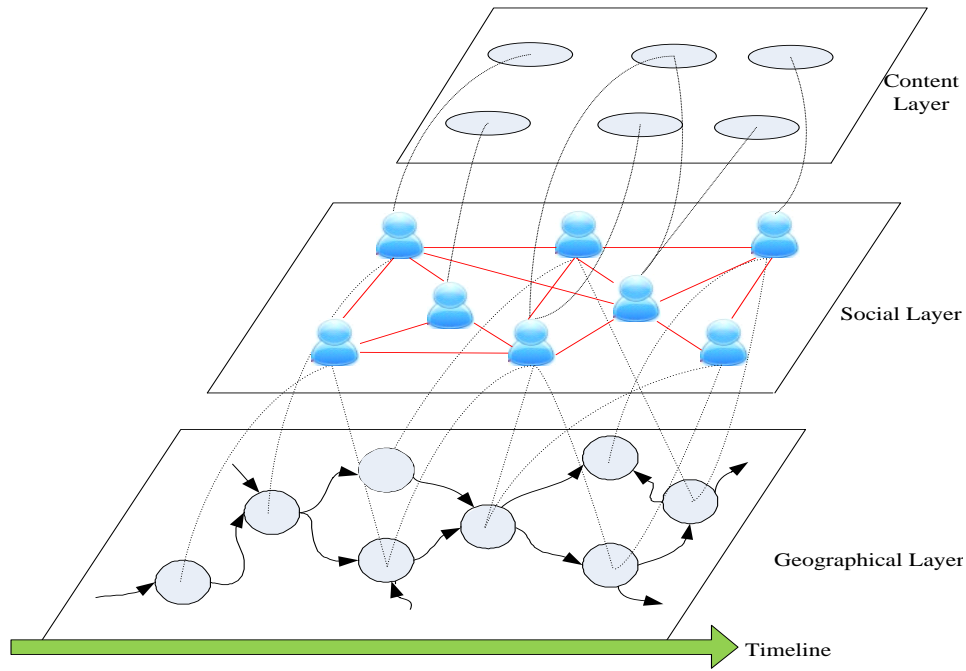

Fig. 1 Framework of LBSNs

(2) Geographical-temporal characteristics

Although geographical locations and tem-poral information are the main components of users' check-ins that are recorded in LBSNs, social activities and social media shared via LBSNs are typically labeled with a location tag. For example, a tourist may share photos with his friends via WeChat (it has become an important social media platform in China, it provides users an innovative way to communi-cate and interact with friends through text messaging, one-to-many messaging, hold-to-talking voice messaging, photo/video sharing, location sharing, and contact information exchange[1]). when he visits Olympic park in Beijing. First, his current geographical location and the time will be recorded by WeChat through his check-ins. Second, the shared photos also indicate this current location by recognizing some of the distinct buildings at Olympic park in Beijing.

(3) Explicit location description

LBSNs both record the longitude and latitude of a location and record additional textual descriptions for a popular POI such as categories, labels, and comments. Therefore, it is easy to distinguish two adjacent stores on a street or two neighboring buildings in a park. This is also why the geographical data in LBSNs is efficiently used in some location-based services (e-commerce recommender, trip planning, and accurate advertising).

(4) Unambiguous social relationships

Like traditional online social networks, LBSNs allow users to add other users as friends, meaning that the social relationships between users are entirely defined by users. The social relationships of all users in LBSNs can be written as a 0-1 matrix, where 1 represents when two users are friends and 0 represents when two users are not friends.

(5) User-driven big data

Users' check-in behavior is user-driven [25] in LBSNs; the user freely decides whether to check-in at a specific location depending on their personal preference. The recent rapid development of location-based services has meant that LBSNs record a large amount of user-generated geographical and social data from billions of users. For example, foursquare (a social-driven location sharing and local search-and-discovery service mobile app[2])had approximately 55 million monthly active users and 10 billion check-ins by December 2016[3]. WeChat had 938 million monthly active users by April 2017 and 639 million users had accessed it on a smartphone each month on average by January 2016[1].

---

[1] https://en.wikipedia.org/wiki/WeChat

[2] https://en.wikipedia.org/wiki/Foursquare
[3] http://expandedramblings.com/index.php/by-the-num

(6) Data sparsity

The increasing use of smart devices and popular LBSNs has meant that the total number of user check-ins in LBSNs has increased. However, the user-driven behaviors for check-in anywhere and anytime has led to significant sparseness in consecutive check-ins on LBSNs. For example, the average number of daily check-ins on Foursquare is 8 million [2] and we can determine that the average number of daily check-ins for each user on Foursquare is: $\frac{8 \times 30}{55} \approx 4.36$.

## 3、Formalization of user modeling for POI recommendations in LBSNs

LBSNs typically consist of a set of $N$ users $U = \{u_i \mid 1 \leq i \leq N\}$ and a set of $M$ POIs $P = \{p_m \mid 1 \leq m \leq M\}$ where each POI belongs to one or more categories. Furthermore, Yelp[2] (a crowd-sourced local business review and social networking site in USA) and Dianping (a crowd-sourced local business review and social networking site in China) also provide the semantics of POIs that contain far more information than just the category. A social networks matrix $S = \{s_{ij} \mid s_{ij} = 0/1, 1 \leq i, j \leq N\}$ represents the social relationships among all users, and $s_{ij} = 1$ indicates the existence of a social relationship between user $u_i$ and $u_j$, whereas $s_{ij} = 0$ indicates no social networks between them, and a user-POI matrix $C = \{c_{ip_m} \mid |c_{ip_m}| \geq 0, 1 \leq i \leq N, 1 \leq m \leq M\}$, $c_{ip_m}$ represents the frequency or comment information of POI $p_m$ visited by user $u_i$.

The goal of user modeling in LBSNs is to learn user's implicit preferences for POIs at the correct time and place; this formation is summarized below.

$$\forall u_i \in U, I^* = \arg\max_{m \in M}(\alpha \mu_{p_m} + \beta \mu(p_m, u_i, l, t))$$

Where $\alpha + \beta = 1$ and $\mu_{p_m}$ is a bias that represents the popularity of $p_m$, $\mu: P \times U \times L \times T$

---

bers-interesting-foursquare-user-stats/
[1]http://expandedramblings.com/index.php/downloads/dmr-wechat-statistics-report/
[2]https://en.wikipedia.org/wiki/Yelp

$\rightarrow R^*$, $L$, $T$ respectively represent users' movement regions and the time.

## 4、The taxonomy of user modeling for POI recommendation in LBSNs

In this section, we briefly review the taxonomy of user modeling in LBSNs according to which type of LBSN data is used in the user modeling approaches for POI recommendation. We divide user modeling algorithms into four categories: pure check-in data-based user modeling, geographical information-based user modeling, spatio-temporal information-based user modeling and geo-social information-based user modeling.

### 4.1 Pure check-in data-based user modeling

User-POI data is usually encoded into a sparse matrix because users only visited a few locations in LBSNs, most elements of the user-location matrix are zero. If a user's demographics and POI categories are added to the user-POI data, the user-POI data is formatted as in Fig. 2.

Since the check-in frequencies recorded in LBSNs implicitly reveal users' preferences for POIs, several studies have adopted a topic model [26], a location hierarchical classification model [27,28], a Latent Dirichlet Allocation [29], a Gaussian kernel approach [30], matrix factorization [31] or a latent factor model [31,32] to infer users' preferences for POI recommendations.

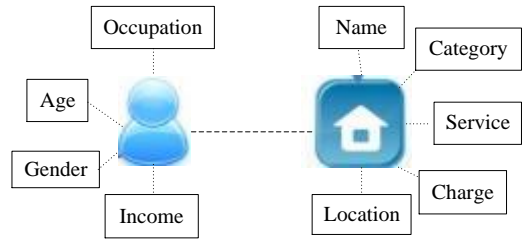

Fig. 2 The Structure of User-POI data

The effectiveness and efficiency when dealing with a large user-item rating matrix $R_{N \times M}$ mean that Matrix factorization techniques [33] have been successfully used in traditional recommender systems. Two low-rank matrices $U_{N \times K}$ and $V_{K \times M}$ ($K \ll \min(N, M)$) are decomposed from the user-item rating matrix $R_{N \times M}$ where $U_{N \times K}$ and $V_{K \times M}$ are treated as a user latent factor and an item latent factor. Matrix factorization techniques can also be employed for POI recommendations in LBSNs. For example, Berjani et al. [34] sough-t to deal with a lack of explicit ratings for POIs in LBSNs by first transforming users' check-ins to ratings information and proposing a regularized matrix factorization-based POI recommendation algorithm.

The objective function was as follows.

$$\min_{U,V} \sum_{(u,i) \in T} (r_{ui} - U_u^T V_i)^2 + \lambda (\|U_u\|^2 + \|V_i\|^2) \quad (1)$$

Where $T$ is the set of user-spot pairs and $\lambda$ is the regularization parameter. Wang et al. [35] proposed a new POI recommendation algorithm to model the importance of venue semantics in user check-in behavior by treating venue semantics as an additional regularizer in the objective optimization func-tion; the objective function is as follows.

$$\min_{U,V} \frac{1}{2} (\|R - V^T U\|_F^2 + \lambda_1 \|U_u\|_F^2 + \lambda_2 \|V_i\|_F^2 + \lambda_3 \sum_{i,j} S_{ij} \|V_i - V_j\|_F^2) \quad (2)$$

Where $\lambda_1, \lambda_2, \lambda_3 > 0$ and $S_{ij} \in [0,1]$ is the semantic similarity between venues $L_i$ and $L_j$.

Moreover, some content information and context information (e.g. POI category, user context, sentiment indication, and timestamp) in LBSNs also reveal different characteristics of users' check-in behaviors. Many research-ers [31,32,36-40] have proposed many conte-xt-aware [36-38] and content-aware [39,40] POI recommendation frameworks in conside-ration of above information.

One major advantage of these approaches is to achieve the purpose of dimensionality reduction and alleviate data sparseness. There is not a standard way to transform users' check-ins into rating data. Another disadvant-age is without considering the geographical, temporal and social influence of users' check-ins.

### 4.2 Geographical information-based user modeling

Like traditional recommender systems, the above-mentioned approaches often treat POIs as items, but do not consider geographical influence, which is a unique characteristic that distinguishes POIs from items in traditional recommender systems. Therefore, leveraging the geographical information of users' check-ins (as shown in Fig.3) can capture the spatial distribution of humans' daily movement and enhance the performance of POI recommendation systems in LBSN.

1. Bayesian model-based user modeling

Similar to Fig. 3, many studies [44-46] have shown that the spatial clustering phenomenon of users' check-ins in LBSNs, which results from users' tendency to visit nearby places rather than distant ones in their daily lives. It is intuitive that the Bayesian model [41] and probabilistic method [42-46] can be employed to model the geographical influence of user check-ins in LBSNs. For example, to model the geographical influence of users' check-in behaviors, Ye et al. [41] utilized the power law distribution to model the geographical influence among POIs and proposed a collaborative POI recommendation algorithm that was based on a naïve Bayesian one. Zhang et al. [43] proposed a probabilistic approach to model personalized geographical influence on user check-in behavior and predict the probability of a user visiting a new location. To model the numbers of centers that are checked-in by different users' LBSNs, Cheng et al. [44] computed the probability of a user checking in to a location via a multi-center Gaussian model and proposed a POI recommendation framework with a combination of user preference, geographical influence, and personalized ranking. Pham et al. [45] proposed an out-of-town region recommendation algorithm in consideration of the spatial influence between POIs to measure a region's attractiveness. By taking the spatial influence of users' check-ins into account, it could narrow the searching space to enhance the performance of recommendation systems. The disadvantage of these approaches is that they could not deal with user cold-start problem.

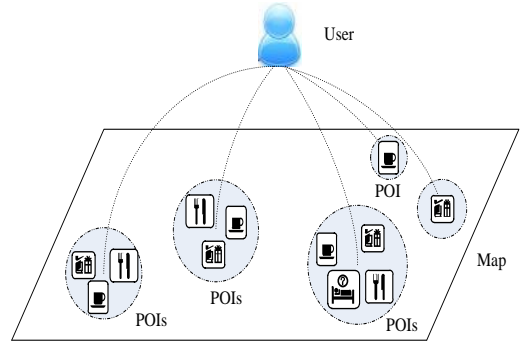

Fig. 3 Spatial cluster of user check-ins

2. Latent factor model

Alongside the development of the matrix factorization technique in the recommendation system, another intuitive method of modeling users based on geographical information in LBSNs is the latent factor model. The main challenge is how to combine the geographical influence of user behaviors with matrix factorization. In general, the inherent spatial feature (e.g. neighbor) of POIs and the spatial clustering phenomenon (e.g. all users who visit POIs tend to cluster together and several geographical regions are automatically formed) are the core geographical influences that are considered in the latent factor model, and are usually treated as additional latent factors in matrix factorization. The state-of-the-art approaches to user modeling can be divided into two groups.

(1) Geographical neighbors

The observations that individuals tend to visit nearby POIs and their geographical neighbors in LBSNs have been effectively used in POI recommendations. For example, Hu et al. [47] proposed a latent factor model for rating predictions that combined the intrinsic characteristics of businesses and the extrinsic characteristics of their geographical neighbors. The predicted rating and objective function were as follows.

$$\hat{r}_{ui} = \mu + b_u + b_i + z +$$
$$\mathbf{p}_u^T (\frac{1}{|R_i|} \sum_{w \in R_i} \mathbf{q}_w + \frac{\alpha_1}{|N_i|} \sum_{n \in N_i} \mathbf{v}_n + \frac{\alpha_2}{|C_i|} \sum_{c \in C_i} \mathbf{d}_c) \quad (3)$$

$$\min_{\mathbf{p}_*,\mathbf{q}_*,b_*,\mathbf{v}_*,\mathbf{d}_*,\beta_*} \sum_{(u,i) \in K} (r_{ui} - \hat{r}_{ui}) + \lambda_1 (\|\mathbf{p}_u\|^2 + \sum_{w \in R_i} \|\mathbf{q}_w\|^2)$$
$$+ \lambda_2 (b_u^2 + b_i^2 + \beta_i^2 + \beta_u^2)$$
$$+ \lambda_3 (\sum_{n \in N_i} \|\mathbf{v}_n\|^2 + \sum_{c \in C_i} \|\mathbf{d}_c\|^2)$$

(4)

Where $\mu$ is the average rating of all known ratings, $b_u$ and $b_i$ represent the user bias and item bias, $\mathbf{p}_u$ represents the latent factors of user $u$, $\mathbf{q}_i$ represents the latent factors of item $i$ for its intrinsic characteristics, $\mathbf{v}_i$ represents the latent factors of item $i$ for its extrinsic characteristics, $\mathbf{d}_c$ represents the latent factors of category $c$, and $\mathbf{q}_w$ represents the latent factors of review word $w$.

Moreover, Li et al. [48] proposed a ranking-based geographical factorization method for POI recommendations that obtains user-preference scores and geographical neigh -bor scores through user-POI matrix factoriza -tion and POI-$k$-nearest neighbor matrix factorization. Feng et al. [49] considered sequential influence, where the next POI is influenced by the current POI within a short period and the geographical influence of a distant POI is less likely to be recommended, and proposed a personalized ranking metric that embeds a model for the next new POI recommendation.

(2) Geographical region

Apart from geographical neighbors, geo -graphical region is another geographical influence that is used in a latent factor model. Many researchers [50-56] have recently discovered spatial clustering phenomena in human mobility behavior and demonstrated its effectiveness in POI recommendations. For example, Liu et al. [54] proposed a novel location recommendation approach that exp -loits instance-level characteristics and region -level characteristics by incorporating two level geographical characteristics into the lear -ning of the latent factors of users and locations. The predicted rating and objective function were:

$$\hat{r}_{ui} = \alpha \mathbf{U}_u \mathbf{L}_i + (1-\alpha) \frac{1}{Z(l_i)} \sum_{l_k \in N(l_i)} Sim(l_i, l_k) \mathbf{U}_u \mathbf{L}_k^T$$

(5)

$$\min_{\mathbf{U},\mathbf{L}} = \frac{1}{2} \sum_{u,i} W_{ui} (r_{ui} - \hat{r}_{ui})^2 + \frac{\lambda_1}{2} \|\mathbf{U}\|_F^2 + \frac{\lambda_2}{2} \|\mathbf{L}\|_F^2 +$$
$$\lambda_3 \sum_{g=1}^{G} \sum_{d=1}^{r} w_g \|\mathbf{L}_{(g)}^d\|_2$$

(6)

Where $\alpha \in [0,1]$ is the instance weighting parameter, $N(l_i)$ is the set of $N$ nearest neighboring locations of $l_i$, $N(l_i) = \sum_{l_k \in N(l_i)} Sim(l_i, l_k)$, $Sim(l_i, l_k)$ is a Gaussian function, $\mathbf{L}_{(g)}^d \in R^{n_g \times 1}$ is the $d^{th}$ column vector in $\mathbf{L}_{(g)}$, and $w_g$ is the weight assigned to $\mathbf{L}_{(g)}^d$.

Furthermore, Liu et al. [55] leveraged a latent region variable to model user mobility behaviors over different activity regions and proposed a geographical probabilistic factor model for POI recommendations. Chen et al. [56] proposed a probabilistic latent model by considering the cluster phenomenon where the users' check-in places were automatically divided into several regions, and how users' psychological behavior could make them prefer a nearby place to a distant one.

The main challenge to latent factor model is to incorporate the geographical information into latent factor and reduce computational complexity.

### 4.3 Spatio-Temporal information-based user modeling

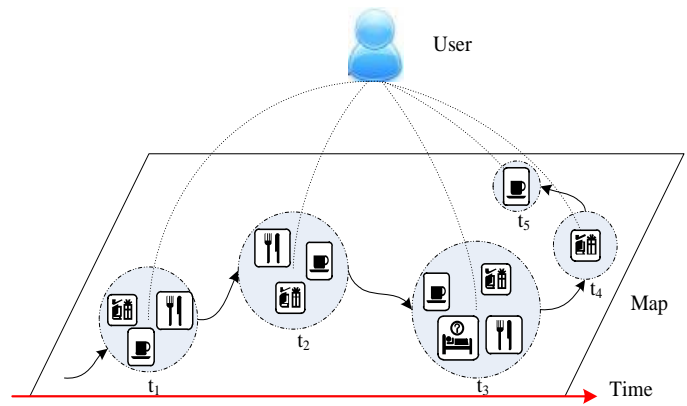

Fig. 4 Check-in sequence in LBSNs

User check-ins demonstrate that short-ranged travels are successive and periodic both spatially and temporally in LBSNs [1, 57,58]. Although users' check-ins exhibit a periodic pattern, which implies users' lifestyle, all POIs visited by users result in check-in sequences, which reveals how two successive POIs can be geographically adjacent and temporally relevant from the perspective of a venue's function (as shown in Fig.4). There-fore, temporal information is an important contextual factor used in user modeling for POI recommendations.

(1) Time-aware user modeling

The time factor is considered a contextual factor and used to enhance the POI recommendation system. The time factor affects human experiences and the temporal clustering phenomenon also exists in our daily lives, not only the geographical clustering phenomenon. For example, most users visit different types of POI at different times in a day and visit different types of POI on weekdays and weekends. For example, they may visit a food-related POI at noon and a nightlife spot in the evening. Most office staff commute from home to their company every weekday morning and shop at a supermarket on weekday afternoons. Some users' temporal POI preferences may be similar, which naturally fits the underlying assumption of collaborative filtering, i.e., users who have similar temporal preferences for certain POIs will likely have similar temporal preferences to others [59,60,64]. User-based recommend-ation methods [61-63,65-67], tensor factorize-tion [64, 68], ranking SVM [69], generative models [70-74], graph-based models [75, 76], and neural networks [77,78] are effective methods for modeling users for POI recomm-endations in LBSNs. For example, Yuan et al. [61] proposed a user-based extended POI recommendation algorithm by leveraging the time factor when computing the similarity between two users and the recommendation score for a new POI. Yao et al. [62] took the compatibility between the time-varying popu-larity of POIs and the regular availability of users into consideration to propose temporal matching between a POI popularity and user regularity recommendation system. Ozsoy et al. [79] proposed a dynamic recommendation algorithm by leveraging users' temporal prefer-ences at different times or days of the week.

These approaches could dynamically produce the recommended POIs in terms of users' temporal preferences. However, the recommended POIs are usually popular with most users and unpopularity POIs (namely, long-tail POIs) and new POIs would not been recommended to any user.

(2) sequential influence-based user mode-ling

Sequential influence is another temporal influence that is utilized in user modeling for POI recommendations in LBSNs. All POIs visited by a user can bring out a check-in sequence, and successive check-ins are typically correlated both spatially and tempo-rally. For example, a user may habitually visit a bar after dinner in a restaurant. This observation reveals that the bar and the restaurant are geographically adjacent and the check-in sequence implies that the temporal relevance from the perspective of venue functions in addition to the user's daily life custom. The Markov chain model is most often exploited to model the sequence pattern for POI recommendations in LBSNs. For example, Cheng et al. [51] took into account two prominent properties in the check-in sequence: personalized Markov chain and region localization and proposed a novel matrix factorization method for POI recommendation, which exploits personalized Markov chain in the check-in sequence and users' movement constraint. He et al. [80] proposed a third-rank tensor with which to model successive check-in behaviors by fusing a personalized Markov chain with a latent pattern. Zhang et al. [81] exploited a dynamic location-location transition graph to model sequential patterns and predicted the probability of a user visiting a location via an additive Markov chain, they also fused seque -ntial influence with geographical influence and social influence into a unified recommen -dation framework. Further -more, they [82] proposed a gravity model weigh the effect of each visited location on the new location, which integrates the spatio-temporal, social and popularity influences by estimating a power-law distribution.

Apart from the Markov chain model, matrix factorization [83], tensor factorization [84], a pairwise ranking model [85, 86], and recurrent neural networks [87] are employed to model the check-in sequential pattern. For example, Chen et al. [84] used a third-rank tensor to compute transitions between catego -ries of users' successive locations and propo -sed a graph-based location recommendation algorithm. Zhao et al. [85, 86] presented two POI recommendation algorithms via a pair -wise ranking model and exploited two differ -ent methods to model the sequential influence from two different aspects. Liu et al. [87] exploited extended recurrent neural networks to model local temporal and spatial contexts and proposed a location recommendation algo -rithm.

Furthermore, Yang et al. [88] used the word2Vec technique to propose a spatio-temp -oral embedding similarity algorithm for location recommendations by treating the time, location, and venue functions of check-in records as virtual "words," check-in sequences as "sentences" and the activity of a neighborhood or user as "documents." Liao et al. [89] proposed a location prediction model by utilizing temporal regularity and sequential dependency. Zhu et al. [90] constructed a user model from location trajectory, semantic trajectory, location popularity, and user famili -arity and proposed a semantical pattern mining and preference-aware POI recommend -ation algorithm. Liu et al. [91] developed a low-rank graph construction model to learn static user preferences and dynamic sequential preferences, and thus proposed a POI recomm -endation algorithm.

Obviously, most of above approaches are content-based recommendation techniques, they make fully use of sequential influence of users' check-ins to model users' spatio-temp -oral preferences for successive POIs. Howe -ver, if a user not check-ins often enough, or is a new user, these approaches would not work well.

### 4.4 Geo-Social information-based user modeling.

Besides geographical and temporal influences, social influence is another source of contextual information exported to user modeling for POI recommendations. Check-ins in LBSNs show [1, 24] that users' long-distance travel is influenced by their friends and users are more likely to visit places that have been visited by their friends. In other words, friends tend to share more common interests than non-friends in LBSNs (the geographical and social relationships of users' check-ins in LBSNs are as shown in Fig. 5). Their observations are widely exploited to model users for POI recommendations [2, 92-98]. For example, Hu et al. [92] proposed a Top-N POI recommendation algorithm by leveraging both the social and topic aspects of users' check-ins. Zhang et al. [93] leveraged the location, time and social information to model users and weighted approximately ranked pairwise losses to achieve top-n POI recommendations. Jia et al. [94] defined several features to measure the influence of friends, rank friends by a sequential random walk with a restart in terms of their influence, and utilized a Bayesian model to characterize the dynamics of friends' influence to predict locations. Li et al. [95] focused on the problem of predicting users' social influence on event recommendations in event-based social networks and proposed a hybrid collaborative filtering model by incorporating both event-based and user-based neighbor -hood influences into matrix factorization. Gao et al. [96] presented an event recommend -ation algorithm by fusing social group influences and individual preferences into a Bayesian latent factor model. Zhang et al. [97] proposed a geo-social collaborative filtering model through a combination of user prefer -ence, social influence, and personalized geogr

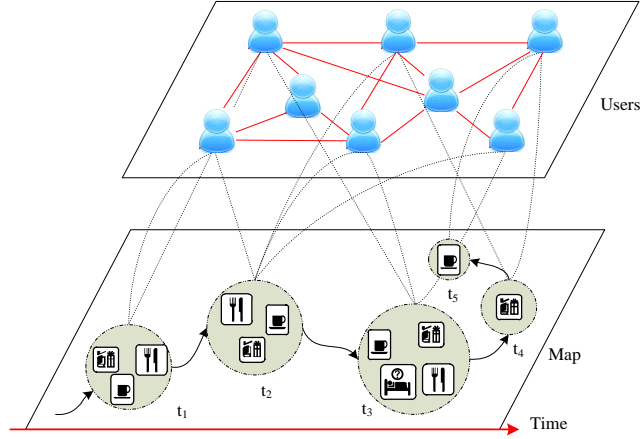

Fig. 5 the spatial and social relationships between users' check-ins in LBSN

-aphical influence that had been learned from users' check-in behaviors by a kernel density estimation approach. Additionally, they proposed a geographical-social-categorical correla-tion enhanced POI recommendation approach in [2] by taking categorical correlations betw-een POIs into consideration. In [98] presented an LDA-based POI recommendation model that jointly mined latent communities, regions, activities, topics, and sentiments from social links between users, venue geographical locations, venue categories, and textual comm-ents on venues.

In addition to traditional recommendation systems (e.g. e-commerce recommendation systems and context-aware recommendation systems), POI recommendation systems in LBSNs also face many challenging problems such as the issue of data sparsity and the user/POI cold-start problem. Incorporating social network ties into certain mathematical models (e.g. matrix factorization, graph model) is an effective solution to cope with such challenges [99-103]. For example, Zhang et al. [100] proposed a local event recommendation approach that took Bayesian Poisson factoriz-ation as its basic unit to model events, social relations, and content text, and handled the cold-start problem by incorporating event textual content and location information into these basic units. Yin et al. [101] proposed a POI recommendation algorithm that was based on a probabilistic generative model, which considered the phenomenon of user interest drift across geographical regions, exploited social and spatial information to enhance the inference of region-dependent personal interests, and alleviated the issue of data sparsity. Yao et al. [37] presented a collaborative filtering POI recommendation method based on non-negative tensor factorize-ation and fused users' social relations as regularization terms of the factorization to improve the recommendation accuracy. Ren et al. [102] exploited a weighted product of user latent factors and POI factors by incorporating a topic with geographical, social, and catego-rical information to enhance the performance of a probabilistic matrix factorization. POI recommendation accuracy can be improved and cold-start problems can be addressed by leveraging the information of friends; Li et al. [103] first defined three types of friend (social friends, location friends, and neighboring friends), and incorporated the set of locations that received individual likes and were checked-in by individual's friends into matrix factorization. The predicted formulation and objective function were as follows.

$$\hat{p}_{ui} = (q_{uc_i} + \varepsilon)\mathbf{U}_u^T \mathbf{V}_i \qquad (7)$$

$$\arg\min_{\mathbf{U},\mathbf{V},\mathbf{Q}} = \sum_u E_u(p_{ui}, p_{uk}, p_{uh}, \hat{p}_{ui}, \hat{p}_{uk}, \hat{p}_{uh})$$
$$+ \frac{\lambda_u}{2}\|\mathbf{U}\|_2^2 + \frac{\lambda_v}{2}\|\mathbf{V}\|_2^2 + \frac{\lambda_q}{2}\|\mathbf{Q}\|_2^2$$
$$\forall i \in M_u^o, \forall k \in M_u^p, \forall h \in M_u^w$$
$$(8)$$

Where $c_i \in \mathbf{Q}$ is the category of location $i$ and $\varepsilon$ is a tuning parameter. For each user $u$, $M_u^o$ represents observed locations, $M_u^p$ potential locations, $M_u^w$ other unobserved locations, and $E_u(\cdot)$ is the loss functions for the observed, potential, and unobserved prefer-ences of user $u$ for a location. Moreover, Guo et. al.[104] exploited the geographical, social information and aspects extracted from user reviews to better model user preferences, then constructed a novel heterogeneous graph by fusing three types of nodes (users, POIs and aspects) and various relations among them, finally transformed the personalized POI reco-mmendation as a graph node ranking proble-m.

Table 1 Statistics on the literature

| Name | 2011 | 2012 | 2013 | 2014 | 2015 | 2016 | 2017 |
|---|---|---|---|---|---|---|---|
| **Conference** | | | | | | | |
| AAAI | | 1 | | | 1 | 5 | |
| IJCAI | | | 1 | | 2 | | |
| WWW | | | | | | | 1 |
| KDD | 2 | 1 | 2 | 1 | 3 | 2 | |
| SIGIR | 1 | | 1 | 1 | 3 | | |
| CIKM | | | 2 | 2 | 1 | 3 | |
| ICDM | | | 1 | 1 | | 1 | |
| DASFAA | | | | | | 2 | 2 |
| ASONAM | | | | | 1 | 1 | |
| Urbcom | | 1 | | | | | |
| SIGSPATIAL | | | | 1 | | | |
| ICONIP | | | 1 | | | 1 | |
| RecSys | | | 1 | | | | |
| PAKDD | | | | | | | 4 |
| **Journal** | | | | | | | |
| TIST | | | | | 2 | 4 | 1 |
| TOIS | | | | 1 | | 1 | |
| TKDD | | | | | | 1 | 1 |
| TKDE | | | | | 1 | 2 | |
| TON | 1 | | | | | | |
| TSC | | | | | 1 | 2 | |
| TCC | | | | | | 1 | |
| TMM | | | | | 2 | | |
| Information Science | | | | | 1 | | |
| Neurocomputing | | | | | | 1 | 3 |
| Knowledge-based systems | | | | | | | |
| **Total** | 4 | 3 | 9 | 7 | 18 | 27 | 12 |

Table 2 Taxonomy by methodology

| categories | | | literatures |
|---|---|---|---|
| context-aware techniques | | | [36-38,47,48,54,59,60,62,64,79] |
| content-based techniques | | | [26-30,39,40,44,45,46] |
| CF | memory-based CF | UCF | [2,59-63,65-67,92-98] |
| | | ICF | [88-91] |
| | model-based CF | | [31,32,34,35,47-56,64,68-78,99-103] |
| hybrid techniques | | | [47,54,81,82] |

These approaches could solve user cold-start problem and alleviate the sparseness of users' check-ins by leveraging social informa-tion in LBSNs. The major challenge is how to incorporate users' social relations into the pop-ular models (such as matrix factorization).

# 5、Statistics on the literature

In this section, we first give some brief statistics on the literature published on well-known journals and conferences in recent year-s, as shown as in Table 1. Inspired by literature [105], we further classify user modeling on POI recommendation into three main categories: context-aware techniques, content-based techniques, collaborative filtering (CF) and hybrid techniques, the CF techniques is composed of memory-based CF and model-based CF, the memory-based CF methods include user-based CF (UCF) and item-based CF(ICF). Finally, we simply list them in Table 2.

# 6、The challenges and new directions in the future

POI recommendation system in LBSNs not only satisfies the basic functions of traditional recommendation systems, but also has the characteristics of location-based servic-es and mobile urban computing. In recent years, many researches have been done in user modeling for POI recommendation in LBSNS. But there are some challenges and new direc-tions that would attract lots of researchers' attention in the future.

(1) Mining users' check-ins and social activities in LBSNs

Users' check-ins and social activities reco-rded in LBSNs usually hide their persona-lized preferences for POIs in real word. To some extent, the spatio-temporal properties of users' check-ins and social activities are the significant assumptions for user modeling for POI recommendation [41,47,54,56,62,88]. Th-erefore, mining users' check-ins and social activities play an important role in POI recom-mendation in LBSNs. The main research contents are as follows: the spatio-temporal distribution of users' check-ins, the similarity of users' check-in trajectories, users' activities tracking and recognition.

(2) The relevance between users' check-ins and social relationships

In the past, it was difficult in collecting users' spatial data and social relationships through a united platform. LBSNs provide a new way to collect these data and a new perspective to study the relevance users' check-ins and social relationships. Users' check-ins in LBSNs usually uncover users' personal behaviors and users' social relation-ships in LBSNs usually reveal users' social behaviors in real word. Qualitative and quantitative analysis of the relevance can be used as an important heuristic in user modeling for POI recommendations [92-98] and inferring new social links [17-19]. With the wide application of location-based POI recommendations, the research on the relev-ance between users' check-ins and social relationships will gradually attract more and more attention.

(3) The interpretation of recommenda-tions

Effective interpretation and clear presenta-tion can make users to fully understand the recommendations, thus improve users' accept-ance of the recommendations and their stick-iness for recommendation systems. Especia-lly，in the scenario of smart device with a relatively small screen and inconvenient input, a more user-friendly interface and game-oriented interpretation are needed. There are few studies on the interpretation and presenta-tion of POI recommendations in LBSNs, Ho-wever, we think that making users to understand the recommendations is likely to be a hot research point in the future, not only referring to a two-dimensional point or a cont-inuous path in the map.

(4) Scalability

Scalability always seems to be a distur-bing problem in recommendation systems. In user-based collaborative filtering algorithm, the computational complexity of user similarity is $O(n^2 \cdot m)$ ,where $n$ is the num-ber of users, $m$ is the average number of ite-ms rated by each user. With the increases in the numbers users and items, the computa-tional complexity of similarity will sharply increase. At present, the solutions of the problem in user-based collaborative filtering involve dimensionality reduction through factorization model and narrowing the sear-ching space by incorporating users' contex-tual information. The same problem always exists in POI recommendation in LBSNs, it is certainly sure that these above solutions are effective as well. Therefore, with the grow-th of data in volume and dimensionality, desi-gning a high-efficiency model will be a long-term-focused research point in POI recommen-dation systems as well as in E-commerce reco-mmendation systems.

(5) Privacy preservation

Users often puzzle over the privacy issue (e.g. location disclosure, sensitive relationship disclosure) when they checking-in in LBSNs [106]. For example, Gundecha et. al. [107]

found that privacy issue is the most concern factor of users when they using location-sharing services. As a matter of fact, users enjoy lots of location-based services by forwardly sharing their current locations in LBSNs, at the same time, inappropriate disclosure of location information poses threats to their privacy. Nowadays, privacy preservation in LBSNs has attracted people's attentions from academic research [108] to industrial applications [109].

## 7、Conclusion

The increasing use of smart devices and LBSNs has led to millions of user-generated data in recent years, how these data can be utilized to understand human mobile behavior and help users make correct decisions has attracted the interest of many researchers from different domains. In this paper, we have focused on reviewing the taxonomy of user modeling for POI recommendation via data analysis in LBSNs. We have divided user modeling algorithms into four categories according to which type of LBSNs data has been fully utilized in user modeling approach-es for POI recommendation: pure check-in data-based user modeling, geographical infor-mation-based user modeling, spatio-temporal information-based user modeling, and geo-social information-based user modeling. At last, summarizing the existing works, we point out the future challenges and new directions in five possible aspects.

## Acknowledgements

This work is supported by National Science Foundation of China (NO. 61602518) and Open Foundation of Hubei Key Laboratory of Intelligent Geo-Information Processing (No.K LIGIP20 16A06). The author declares that the-re is no conflict of interest regarding the pub-lication of this paper.